\documentclass[sigconf, authorversion, balance]{acmart}

\usepackage[hide]{chato-notes}
\usepackage{graphicx}
\usepackage{placeins}
\usepackage{enumitem}
\usepackage{multirow}
\usepackage{subfig}
\usepackage{makecell}
\usepackage{babel}
\usepackage{svg}
\usepackage{hyperref} 
\usepackage{multirow}
\usepackage[normalem]{ulem}
\useunder{\uline}{\ul}{}
\usepackage{caption}
\usepackage{tabularx}
\usepackage[flushleft]{threeparttable}

\graphicspath{ {./images/} }

\usepackage[para]{footmisc}

\newcommand{\craig}[1]{\textcolor{black}{#1}}%
\newcommand{\sasha}[1]{\textcolor{black}{#1}}%
\newcommand{\crs}[1]{\textcolor{black}{#1}}%
\newcommand{\scr}[1]{\textcolor{black}{#1}}%

\newcommand{\srs}[1]{\textcolor{black}{#1}}

\newcommand{\srm}[1]{\textcolor{black}{#1}}

\newcommand{\sre}[1]{\textcolor{black}{#1}}

\definecolor{brightmaroon}{rgb}{0.76, 0.13, 0.28}

\usepackage{marginnote}

\copyrightyear{2022}
\acmYear{2022}
\setcopyright{acmcopyright}\acmConference[RecSys '22]{Sixteenth ACM Conference on Recommender Systems}{September 18--23, 2022}{Seattle, WA, USA}
\acmBooktitle{Sixteenth ACM Conference on Recommender Systems (RecSys '22), September 18--23, 2022, Seattle, WA, USA}
\acmPrice{15.00}
\acmDOI{10.1145/3523227.3548487}
\acmISBN{978-1-4503-9278-5/22/09}

\begin{document}

\title{A Systematic Review and Replicability Study of BERT4Rec for Sequential Recommendation}

\author{Aleksandr Petrov}
\email{a.petrov.1@research.gla.ac.uk}
\affiliation{
  \institution{University of Glasgow}
  \country{United Kingdom}
}

\author{Craig Macdonald}
\email{Craig.Macdonald@glasgow.ac.uk}
\affiliation{
  \institution{University of Glasgow}
  \country{United Kingdom}
}

\begin{abstract}
BERT4Rec is an effective model for sequential recommendation based on the Transformer architecture. In the original publication, BERT4Rec claimed superiority over other available sequential recommendation approaches (e.g.\ SASRec), and it is now frequently being used as a state-of-the art baseline for sequential recommendations. However, \sre{not all} subsequent publications confirm\sre{ed} this result and proposed other models that were shown to outperform BERT4Rec in effectiveness. In this paper \sasha{we systematically review all publications that compare BERT4Rec with another popular Transformer-based model, namely SASRec, and show that BERT4Rec results are not consistent within these publications. To understand the reasons behind this inconsistency, we}  analyse the available implementations of BERT4Rec and show that \sasha{we fail to reproduce results of the original BERT4Rec publication when using \sasha{their}  default configuration parameters}. \sasha{However, we are able to replicate \sre{the} reported results with the original code if training for a much longer amount of time (up to 30x) compared to the default configuration}. We also propose our own implementation of BERT4Rec based on the Hugging Face Transformers library, which we demonstrate replicates \sre{the originally reported results on} 3 out 4 datasets, \sasha{while requiring up to 95\% less training time to converge}. Overall, from our systematic review and detailed experiments, we conclude that BERT4Rec does indeed exhibit state-of-the-art effectiveness for sequential recommendation, but only when trained for a sufficient amount of time. \sasha{Additionally, we show that our implementation can further benefit from adapting other Transformer \sasha{architectures} that are available in the Hugging Face Transformers library (e.g.\ using disentangled attention, as provided by DeBERTa, or larger hidden layer size cf. ALBERT).}
 \sasha{For example, on the MovieLens-1M dataset\sre{,} we demonstrate that both these models can improve BERT4Rec performance by up to 9\%. Moreover, we show that an ALBERT-based \craig{BERT4Rec} model achieves better performance on that dataset than state-of-the-art results reported in the most recent publications.}

\end{abstract}

\settopmatter{printfolios=true}

\maketitle

\section{Introduction}

\emph{Sequential Recommender Systems} \sasha{are a class of recommender systems that use} \sre{the} order of interactions \sasha{to make recommendations}. The goal of such recommendation models is to predict the next user-item interaction by the sequence of user's past interactions. This goal is also known as \emph{next item prediction task}. While early sequential recommender systems applied Markov Chains~\cite{rendle2010fpmc}, more recently neural networks based models have been shown to outperform traditional models~\cite{hidasi2015gru4rec, tang2018caser, yuan2019nextitnet, ma2019hierarchical}.  Since \sre{the} arrival of the Transformer neural architecture~\cite{vaswani2017attention} and, in particular the BERT~\cite{devlin2018bert}
 language model, Transformer-based sequential recommendations models, such as SASRec~\cite{kang2018sasrec}, S\textsuperscript{3}Rec~\cite{zhou2020s3}, \sre{LightSANs~\cite{fan2021lighter},  NOVA-BERT~\cite{liu2021noninvasive} and  DuoRec~\cite{qiu2022contrastive} }have achieved state-of-the-art performance in next item prediction.

\begin{figure}
  \includesvg[width=\linewidth]{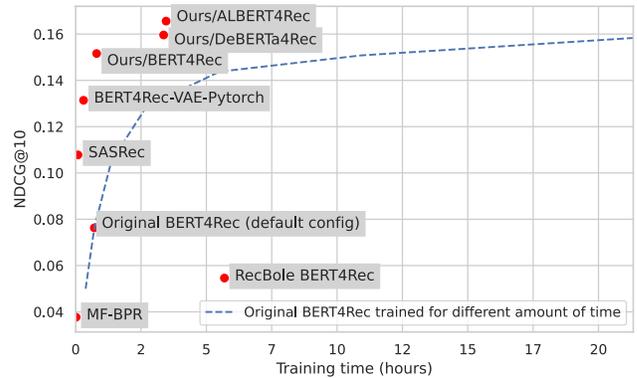}
  \caption{Comparison of four implementations of BERT4Rec (Original, RecBole, BERT4Rec-VAE-Pytorch, Ours/BERT4Rec) with two baseline models (MF-BPR, SASRec) and two more advanced Transformer-based models (Ours/ALBERT4Rec, Ours/DeBERTa4Rec) on the MovieLens-1M dataset. All points represent models with their default configuration.  The dashed line represents NDCG@10 of the original BERT4Rec implementation trained for different amounts of training time.}%
\label{fig:main}
\end{figure}

\sre{In particular,} BERT4Rec is \craig{a highly-cited Transformer-based} \craig{sequence recommendation} model proposed by Sun et al.~\cite{sun2019bert4rec}, which adapts the BERT language model, by equating tokens with items and sentences with sequences of item interactions. \sasha{\scr{Sun et al.}~\cite{sun2019bert4rec} argued that}, one of the main advantages of BERT4Rec compared to other Transformer-based models, such as SASRec, is that it uses an \sasha{item masking training task}~\cite{devlin2018bert} -- \scr{this argument was further investigated and confirmed by Petrov and Macdonald~\cite{petrov2022effective}}. Applied to recommendation, the main idea of this training task is to replace random items in the training sequences with a special \emph{\sre{[mask]}} token and train the model to recover these masked tokens; we describe details of this task in Section~\ref{ssec:bertrec_description}. 

\looseness -1 In the original publication~\cite{sun2019bert4rec}, BERT4Rec \sre{was} claimed to achieve significant superiority over existing neural and traditional approaches, however subsequent publications \sasha{by a number of different authors} (e.g.~\cite{zhou2020s3, yue2021black, fan2021continuous}) did not confirm this superiority. Our goal therefore is to understand the reasons behind this discrepancy. In particular, we analyse \scr{three open source} implementations of BERT4Rec with an aim to reproduce the originally reported results using these implementations. \craig{We find} \sasha{that there is a \sre{marked} discrepancy in both effectiveness and efficiency of these implementations. We also demonstrate that training original implementation of BERT4Rec with its default configuration results in an underfitted model, and it requires up to 30x more training time in order to replicate \sre{the} originally reported results.} 
 
Ultimately, we show that, with proper configuration, BERT4Rec achieves better performance than earlier models such as SASRec, and that some results reported in subsequent papers are based on underfitted versions of BERT4Rec. Moreover, we show that \srs{an appropriately trained BERT4Rec can match or outperform later models (e.g.\ DuoRec~\cite{qiu2022contrastive}, LightSANs~\cite{fan2021lighter} \& NOVA-BERT~\cite{liu2021noninvasive}) and therefore may still be used as a state-of-the-art sequential recommendation model}. 

\sasha{In addition, we propose our own implementation of BERT4Rec that is based on a popular Hugging Face Transformers library~\cite{wolf2019huggingface}. \srs{Hugging Face Transformers is a popular machine learning library with a large supporting community. For instance, its GitHub repository has almost 62K stars and 15K forks,\footnote{\href{https://github.com/huggingface/transformers}{https://github.com/huggingface/transformers}} which makes it the second most popular machine learning library on GitHub.\footnote{After Tensorflow, according to \href{https://github.com/EvanLi/Github-Ranking}{https://github.com/EvanLi/Github-Ranking}} It has became a de facto standard for publishing Transformer-based models \sre{and therefore} contains well optimised and efficient versions of Transformer-based models. Based on this and the fact that it was already successfully applied for recommendations~\cite{de2021transformers4rec},  we expect that using it as a backbone for building Transformer-based recommenders should be} both effective and efficient. Indeed, we show that our implementation with default configuration replicates the results published by \sre{Sun et al.}~\cite{sun2019bert4rec} and requires up to 95\% less time to achieve these results. Moreover, the BERT architecture in our implementation can be easily replaced with other models available in the Hugging Face Transformers library. We demonstrate \scr{by} using two examples (DeBERTa~\cite{he2020deberta} and ALBERT~\cite{lan2019albert}) that such replacement can lead to further improved effectiveness.}

\sasha{Figure~\ref{fig:main} summarises the results of our findings on the MovieLens-1M dataset. \srs{Each point on the figure corresponds to a default configuration of a recommendation model and the dashed line represents performance of the original BERT4Rec code when trained for different amount\sre{s} of time.}\dnote{what are the points vs lines -- done} \srs{The figure} shows that depending on the chosen BERT4Rec implementation, \craig{the observed NDCG@10 effectiveness can} vary from 0.0546 for RecBole implementation to 0.156 (our implementation, 3x difference). It also shows that the required training time also depends on the implementation. For example, \sre{the} BERT4Rec-VAE implementation requires \srs{18 minutes} to train with default configuration whereas \srs{the} original implementation requires \srs{44 minutes} while achieving poorer NDCG@10 \sre{than BERT4Rec-VAE}. The figure also show that the original implementation can reach the same level of performance as the best implementation, however it requires much more training compared to what is specified in the default \sre{configuration}.  
The figure also \craig{portrays} results \sre{of our implementation based on} \sre{DeBERTa~\cite{he2020deberta} and ALBERT~\cite{lan2019albert} Transformers} and shows that these models can outperform BERT4Rec, however they require more time than our BERT4Rec implementation. \sre{Nevertheless}, \sre{their training times are still} 85\% smaller than what is required for the \sre{full} convergence of \sre{the} original BERT4Rec \sre{implementation}.}

\srs{In short, the contributions of the paper are: (1) a systematic review of the papers comparing BERT4Rec and SASRec, \sre{which} shows that the results of such comparisons are not consistent; (2) an analysis of \sre{the} available implementations of BERT4Rec, \sre{which} shows that frequently these implementations fail to reproduce results reported in~\cite{sun2019bert4rec} when trained with \sre{their} default parameters, (3) an analysis of impact of the training time to recommendation performance of the original BERT4Rec implementation; (4) a new Hugging Face based implementation of BERT4Rec that successfully replicates the results reported in~\cite{sun2019bert4rec}; (5) two new recommendation models: DeBERTa4Rec and ALBERT4Rec that improve quality of our implementation via replacing BERT model with DeBERTa and ALBERT respectively}. The code and the documentation for this paper \srs{can be found} in the virtual appendix in our GitHub repository.\footnote{\href{https://github.com/asash/bert4rec_repro}{\scr{https://github.com/asash/bert4rec\_repro}}}

\srs{The rest of the paper is organised as follows: Section~\ref{sec:background} formally introduces \sre{the }sequential recommendation problem and the main approaches to solve it; section~\ref{sec:SASRecAndBert4Rec} describes details of  SASRec and BERT4Rec, popular Transformer-based sequential recommendations models; Section~\ref{sec:systematic_review} contains a systematic review of papers comparing SASRec and BERT4Rec; Section~\ref{sec:implementations} describes available BERT4Rec implementations; Sections~\ref{sec:experiments:setup} and~\ref{sec:experiments:results} describe experimental setup and the results of the experiments respectively; Section~\ref{sec:discussion} put this work in context, wrt. related work and performances observed in recent publications; Section~\ref{sec:conclusion} summarises the results and contains final remarks.}

\section{Background}
\label{sec:background}
The goal of the sequential recommendation is to predict \craig{the} next interaction in a sequence. Consider a set of users $u \in U$, \craig{where} each user has \craig{made} a sequence of interactions $s \in S = \{i_1, i_2, i_3 ... i_t\}$, \craig{and} $i_k \in I$ denotes items. The task of the recommender system is to predict next element $i_{t+1}$ in the sequence. More formally, the goal of the sequential recommender system $M$ is to produce a list of $K$ items ranked by the probability of being the sequence continuation: 
\[
           M(s) \rightarrow \{i_{s_1}, i_{s_2}, i_{s_3}, i_{s_4} ...\ i_{s_{K}}\}
\]

Some of the first sequential models were based on Markov Chains~\cite{rendle2010fpmc}, however over the last few years the best models for sequential recommendations are based on neural networks. Early neural network-based architectures for sequential recommendations included recurrent neural networks~\cite{hidasi2015gru4rec, hidasi2018gru4recplus} and \craig{convolutional} neural networks~\cite{tang2018caser, yuan2019nextitnet}, however,  following the advancements in natural language processing, more recent models \sre{have} adopted \sre{the} self-attention mechanism~\cite{li2017narm, liu2018stamp} and \craig{the} Transformer architecture~\cite{sun2019bert4rec, zhou2020s3, kang2018sasrec, petrov2021booking}. Transformer-based models consistently outperform both traditional (non-neural) models, as well as earlier neural architectures~\cite{sun2019bert4rec, kang2018sasrec, zhou2020s3}.

Two popular Transformer-based models are SASRec~\cite{kang2018sasrec} and BERT4rec~\cite{sun2019bert4rec}. The main difference between these two models is the training task: during training SASRec only uses past interactions to predict future ones, whereas BERT4Rec adopts items masking, which allows to use both past and future interactions to recover masked items. We describe the details of these models and their training tasks in Section~\ref{sec:SASRecAndBert4Rec}. BERT4Rec was published one year later than SASRec, and in the original publication it uses SASRec as a baseline. In the original publication, the authors reported improvements over SASRec on all four datasets they used for comparison. Indeed, for example by the NDCG@10 metric, the authors reported 14.02\% improvement on the Beauty dataset, 5.31\% on the Steam dataset, 10.32\% on the MovieLens-1M dataset,  and  14.47\% on the MovieLens-20M dataset. However, these results were not consistently confirmed in \sre{some subsequent} publications: for example \sre{Zhou et al.}~\cite{zhou2020s3} reported -5.73\% NDCG@10 decrease on Beauty dataset, \sre{Qiu et al.}~\cite{qiu2022contrastive} reported -47.01\% on  MovieLens-1M and \sre{Yue et al.}~\cite{yue2021black} reported -0.8\% on Steam. This paper analyses the reasons for this discrepancy. In the next section, we describe the architectures of the SASRec and BERT4rec models. 

\section{SASRec and BERT4Rec}
\label{sec:SASRecAndBert4Rec}

\subsection{Transformer}
Both SASRec and BERT4Rec are based on the Transformer~\cite{vaswani2017attention} architecture, \sre{which was}, originally designed for natural language processing. The original Transformer architecture consists of the encoder and decoder parts\inote{change word, something more formal?}, however \sre{recommendation} models use only the encoder part of a \sre{Transformer}. 

\begin{figure*}
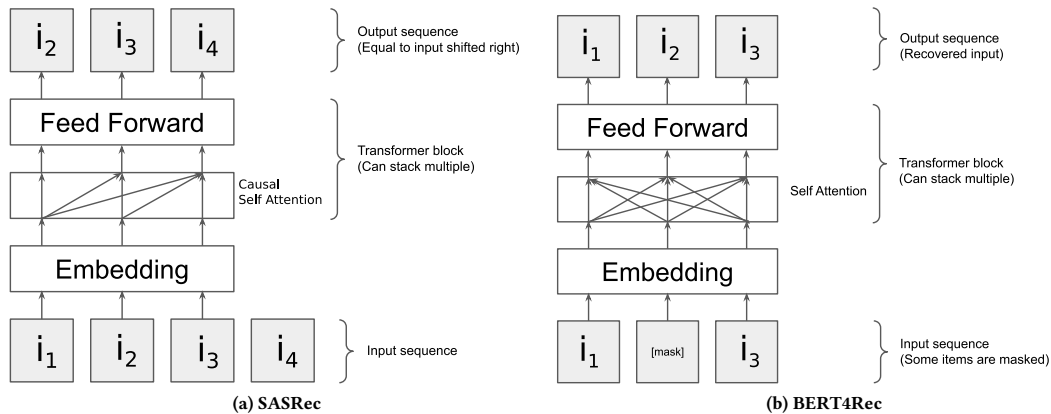

    \centering
    \subfloat[SASRec]{
        \label{subfig:sasrec}
        \includesvg[width=0.4\linewidth]{SASRec_arch.svg}
    }
    \subfloat[BERT4Rec]{
        \label{subfig:bert4rec}
        \includesvg[width=0.4\linewidth]{BERT4Rec_arch.svg}
    }
  \caption{
      Architectures of SASRec \& BERT4Rec. SASRec uses casual (unidirectional) Self Attention, whereas BERT4rec uses regular (bidirectional) Self Attention. SASRec aims to predict the input sequence shifted right, whereas BERT4Rec recovers masked items.
  }\label{fig:modelsarch}
\end{figure*}

Essentially, a Transformer encoder is a sequence-to-sequence model, that translates a sequence of tokens to a sequence of vectors.  It encodes the sequence of the input tokens using an embedding layer, sums these embeddings with positional embeddings (vectors encoding position of the token in the sequence), and passes them through several Transformer {\em blocks}.  \craig{A} Transformer block is the core component of the architecture that consists of a multi-head self attention layer, a pointwise feed-forward layer and residual connections. We refer to the original publication~\cite{vaswani2017attention} for more details about the architecture. In order to use this architecture, SASRec and BERT4Rec use sequences of \sre{item} ids as the inputs. Each input sequence represents a history of interactions of a single user. An output of the Transformer encoder is a sequence of vectors. Both SASRec and BERT4Rec project each vector from the encoder output sequence to the score distribution over items via multiplying these vectors by the matrix of the learnable item embeddings. Figure \ref{fig:modelsarch} illustrates architectures of both models. We now describe details of the two models and highlight their key differences. 

\subsection{SASRec}
\label{ssec:sasrec}
\sasha{Figure~\ref{subfig:sasrec} illustrates the SASRec model}. \sasha{The model} uses a sequence of items (a history of interactions of a single user) as its input and predicts the same sequence shifted by one element to the right. This means, that the last element in the predicted sequence corresponds to the future next iteration, which exactly matches next item prediction - the main task of the sequential recommendation. Indeed, at the inference time, SASRec only uses the last element in the predicted sequence. 

SASRec uses \craig{the} \sre{causal} (unidirectional) version of self attention: to predict  element $i_k$ of the output sequence it can only \sre{access} elements $1, 2 .... i_k$ from the input sequence. Casual self attention prevents the model from experiencing information leakage: with regular (bidirectional) self attention, the task would be trivial - the model would just copy element $i_{k+1}$ from the input sequence as its $k^{th}$ output. 

\subsection{BERT4Rec}\label{ssec:bertrec_description}
\sasha{Figure~\ref{subfig:bert4rec} illustrates BERT4Rec architecture}. As we can see, it is very similar to the SASRec architecture. One important difference is that BERT4Rec uses regular (bidirectional) self attention instead of casual attention. BERT4Rec also uses different training task - instead of predicting \sre{the} shifted sequence, it employs \sre{an} item masking training task. At the training time, some of the items in the input sequences are replaced with \sre{a} special \emph{[mask]} token. The goal of the model is to recover these masked tokens. Items masking task allows to generate more training samples - up to ${n \choose k}$, where $n$ is the sequence length and $k$ is the number of masked items - out of a single sequence. At the inference time, BERT4Rec adds \sre{a} \emph{[mask]} token to the end of the sequence of interactions and then \sasha{for each item in the catalogue BERT4Rec computes probability of being this {\em [mask]} token.} 

Items masking is loosely connected to the next item prediction task, which makes it harder to train. However, as shown in the original BERT4Rec publication~\cite{sun2019bert4rec}, its bidirectional nature may be advantageous and improve the quality of the model. Nevertheless, as we show in the next section, the question whether or not BERT4Rec is a superior model compared to SASRec is still an open question and recent publications do not give a definitive answer to this question.

\section{Systematic review of SASRec and BERT4Rec performance}
\label{sec:systematic_review}
The results published in the original BERT4Rec paper~\cite{sun2019bert4rec} are indeed promising. According to the publication, BERT4Rec  outperformed the second best model, SASRec, on four different datasets (Beauty~\cite{he2016ups}, Steam~\cite{kang2018sasrec}, MovieLens-1M~\cite{harper2015movielens} \sre{and}  MovieLens-20M~\cite{harper2015movielens}) and across six different evaluation metrics (Recall@\{1, 5, 10\}, NDCG@\{5, 10\} and MRR). In order to validate this result and estimate its reproducibility, we perform a systematic review of the papers that compare SASRec and BERT4Rec. We formulate two hypotheses that we test in our systematic review: 

\todo{these labels are broken}
\begin{enumerate}[font={\bfseries}, label={H\arabic*}]
    \item
        \label{hyp:non_consistent}
        Overall, BERT4Rec is not systematically better than SASRec in the published literature. 
    \item 
        \label{hyp:poor_repro}
        Even when experimental setup is similar (e.g.\ experiment on the same datasets), the outcome of the comparison of BERT4Rec and SASRec may be different, which indicates poor replicability of BERT4Rec results. 
\end{enumerate}

Hypothesis~\ref{hyp:non_consistent} addresses  a general question if BERT4Rec is indeed a superior model when compared to SASRec; i.e.\ it outperforms SASRec across a large number of different datasets and metrics. 
Hypothesis~\ref{hyp:poor_repro} addresses the question of reproducibility; we look in particular datasets and metrics and check if the experiment results are consistent within this experimental setup. 

\subsection{Review Methodology}
To test Hypotheses~\ref{hyp:non_consistent} and~\ref{hyp:poor_repro} we conducted a review of all papers citing the original BERT4Rec publication~\cite{sun2019bert4rec} according to the Google Scholar website.\footnote{\url{http://scholar.gooogle.com}}
From these papers, we chose to include only those papers that used both BERT4Rec and SASRec models for their experiments, and examined their observed performances. We further excluded papers that were not peer reviewed to minimise chances of relying on non-verified experiments.\footnote{Other examples of excluded publications include a paper withdrawn by authors -- and only available in the web archive -- and an MSc student thesis.}

\looseness -1 We considered performing a {\em meta analysis}, by aggregating improvements across multiple papers (e.g.\ ``average improvement by BERT4Rec in NDCG@10 over SASRec"). However, we \sre{observe} that the \sre{identified} papers applied \sre{various}  evaluation measures (e.g.\ varying rank cutoffs) and methodology (e.g.\ different negatives sampling strategies, see Section~\ref{ssec:metrics}). Instead we considered three outcomes of the experiments: "BERT4Rec wins", "SASRec wins" and "Tie", and \sre{relied} on counting. In particular, we counted that a particular model wins a comparison if it was better according to all metrics used in the experiment. If a model \sre{was} better according to one subset of metrics and worse according to another, we counted a tie. 

To test \sre{the} overall superiority of BERT4Rec over SASRec (Hypothesis~\ref{hyp:non_consistent}), we determine the total numbers of each possible outcome (BERT4Rec wins; SASRec wins; Tie) and check whether or not there is any considerable amount of situations when SASRec wins over BERT4Rec. To test results reproducibility on different datasets (Hypothesis~\ref{hyp:poor_repro}), we aggregate the results by datasets and count experiment outcomes for each dataset independently.

\subsection{Systematic Review Results}
We reviewed the 37\sre{0} papers citing BERT4Rec according to Google Scholar on 25/03/2022.\footnote{The spreadsheet of analysed papers is also included in \href{https://github.com/asash/bert4rec_repro}{our virtual appendix}.} Following the inclusion criteria (comparison with SASRec) we found 58 publications containing such comparisons, and thereafter excluded 18 (not peer-reviewed etc.).  This left a total of 40 publications that compare BERT4Rec and SASRec, making a total with 134 total comparisons on 46 different datasets (3.35 comparisons per paper on average, 2.91 comparisons per dataset on average).

\subsubsection{\ref{hyp:non_consistent}. Overall results consistency}

Table~\ref{table:systematic_review} summarises the results of  \sasha{our systematic review}. As we can see from the \sasha{"Total" row of the table}, BERT4Rec indeed wins \sasha{the} majority of the comparisons (86 out of 134, 64\%). However, the numbers of comparisons, which SASRec won (32 out of 134, 24\%) or at least achieved a tie (16 out of 134, 12\%) is not negligible: 36\% overall. Therefore we can conclude that BERT4Rec is not consistently superior compared to SASRec in the published literature, which validates Hypothesis~\ref{hyp:non_consistent} \sre{and not consistent with the original BERT4Rec paper.}\inote{and not consistent with original bert4rec paper?} We now investigate if this inconsistency can be explained by the differences in datasets, i.e.\ if the reason is that BERT4Rec is better on some datasets, but SASRec is better on others. 

\subsubsection{\ref{hyp:poor_repro}. Poor replicability of BERT4Rec} \sasha{We now turn again to} Table~\ref{table:systematic_review} \sasha{and analyse} the results of the systematic review aggregated by dataset. The table includes all datasets from our review \sasha{appearing in at least 5 papers}. From the table, we observe that the proportion of outcomes is indeed highly dependent on the dataset. For example, on the both ML-20M and Steam datasets, BERT4Rec won 7 out 8 comparisons (88\%), whereas on Sport it won just 1 out of 6 (17\%) and on Toys it won 0 out of 5 (0\%) comparisons. Therefore, it appears that the disparity in the overall results can be explained by \craig{some} salient characteristics of \craig{the} datasets.

However there is still some level of inconsistencies within datasets. For example, on the two most popular datasets, Beauty and ML-1M, the proportion of  experiments which BERT4Rec did not win roughly matches the overall result: 7 out of 19 for Beauty (37\%) and 5 out of 13 for ML-1M (28\%). Importantly, the original BERT4Rec paper~\cite{sun2019bert4rec} found that BERT4Rec was superior by a statistically significant margin on both of these datasets, and as we can see a large number of papers failed to replicate this result, which validates our Hypothesis~\ref{hyp:poor_repro}. 

\srs{\crs{Overall, we argue that the lack of agreement in the observed results} is problematic: it means that papers have used different versions of BERT4Rec with different configurations and ultimately many of these papers may be using poor configurations of BERT4Rec as their baselines. This leads us to investigate the available open source implementations of BERT4Rec and their hyperparameter settings.}

\begin{table*}[tb]

\caption{Results of BERT4Rec vs SASRec comparisons in the peer-reviewed publications. Bold denotes model with more wins on a dataset. Asterisk (*) denotes datasets used in the original BERT4Rec paper~\cite{sun2019bert4rec}. Only the datasets with \sasha{appearing in at least 5 papers are presented (8 datasets out of 46), however the comparison results on all other 38 datasets are included in the "Total" numbers.}}\label{table:systematic_review}
\begin{tabular}{l|r|lll|lll}
\toprule
                                dataset &  \makecell{total} &    \makecell[r]{BERT4Rec\\wins} &      \makecell[r]{SASRec\\wins} &    \makecell[r]{Ties} &                                                                                                                                                                                                                                                                                                                                      \makecell[r]{BERT4Rec\\wins papers} &                                                                                                               \makecell[r]{SASRec\\wins papers} &                                              \makecell[r]{Ties\\papers} \\
\midrule
                Beauty*~\cite{he2016ups} &                19 & \makecell[r]{\textbf{12 (63\%)}} &           \makecell[r]{5 (26\%)} & \makecell[r]{2 (11\%)} &                                       \makecell[r]{\cite{sun2019bert4rec}, \cite{ma2020disentangled}, \cite{zhou2021contrastive}, \cite{liu2021augmenting},  \\ \cite{li2021intention}, \cite{cho2020meantime}, \cite{zhang2020match4rec}, \cite{tong2021pattern},  \\ \cite{yue2021black}, \cite{he2021locker}, \cite{wu2021seq2bubbles}, \cite{sun2022sequential} \\ } & \makecell[r]{\cite{zhou2020s3}, \cite{li2021lightweight}, \cite{zhang2021behavioral},  \\ \cite{wang2022sequential}, \cite{qiu2022contrastive}} & \makecell[r]{\cite{bian2021contrastive}, \cite{dallmann2021case}}\\ 
 \midrule 

       ML-1M*~\cite{harper2015movielens} &                18 & \makecell[r]{\textbf{13 (72\%)}} &           \makecell[r]{3 (17\%)} & \makecell[r]{2 (11\%)} & \makecell[r]{\cite{sun2019bert4rec}, \cite{ma2020disentangled}, \cite{zhou2021contrastive}, \cite{li2021intention},  \\ \cite{cho2020meantime}, \cite{kang2021entangled}, \cite{tong2021pattern}, \cite{he2021locker},  \\ \cite{wu2021seq2bubbles}, \cite{song2021capturing}, \cite{padungkiatwattana2022arerec}, \cite{potter2022gru4recbe}, \cite{li2022recguru}} &                                                         \makecell[r]{\cite{fan2021lighter}, \cite{yue2021black}, \cite{qiu2022contrastive} \\ } &  \makecell[r]{\cite{zhang2020match4rec}, \cite{dallmann2021case}}\\ 
 \midrule 

             Yelp~\cite{asghar2016yelp} &                10 &  \makecell[r]{\textbf{6 (60\%)}} &           \makecell[r]{4 (40\%)} &  \makecell[r]{0 (0\%)} &                                                                                                                                                                                       \makecell[r]{\cite{zhou2020s3}, \cite{amjadi2021katrec}, \cite{bian2021contrastive}, \cite{wang2022sequential},  \\ \cite{qiu2022contrastive}, \cite{padungkiatwattana2022arerec}} &                          \makecell[r]{\cite{fan2021lighter}, \cite{li2021lightweight}, \cite{zhang2021behavioral},  \\ \cite{li2021extracting}} &                                                    \makecell[r]{}\\ 
 \midrule 

      Steam*~\cite{pathak2017generating} &                 8 &  \makecell[r]{\textbf{7 (88\%)}} &           \makecell[r]{1 (12\%)} &  \makecell[r]{0 (0\%)} &                                                                                                                                                                    \makecell[r]{\cite{sun2019bert4rec}, \cite{ma2020disentangled}, \cite{zhou2021contrastive}, \cite{li2021intention},  \\ \cite{zhang2020match4rec}, \cite{dallmann2021case}, \cite{wu2021seq2bubbles}} &                                                                                                               \makecell[r]{\cite{yue2021black}} &                                                    \makecell[r]{}\\ 
 \midrule 

      ML-20M*~\cite{harper2015movielens} &                 8 &  \makecell[r]{\textbf{7 (88\%)}} &            \makecell[r]{0 (0\%)} & \makecell[r]{1 (12\%)} &                                                                                                                                                                    \makecell[r]{\cite{sun2019bert4rec}, \cite{ma2020disentangled}, \cite{zhao2021adversarial}, \cite{li2021intention},  \\ \cite{cho2020meantime}, \cite{wu2021seq2bubbles}, \cite{potter2022gru4recbe}} &                                                                                                                                  \makecell[r]{} &                             \makecell[r]{\cite{dallmann2021case}}\\ 
 \midrule 

                 Sports~\cite{he2016ups} &                 6 &           \makecell[r]{1 (17\%)} &  \makecell[r]{\textbf{4 (67\%)}} & \makecell[r]{1 (17\%)} &                                                                                                                                                                                                                                                                                                                                       \makecell[r]{\cite{li2022recguru}} &                            \makecell[r]{\cite{zhou2020s3}, \cite{li2021lightweight}, \cite{zhang2021behavioral},  \\ \cite{qiu2022contrastive}} &                          \makecell[r]{\cite{bian2021contrastive}}\\ 
 \midrule 

LastFM~\cite{cantador2011heterogeneity} &                 6 &  \makecell[r]{\textbf{4 (67\%)}} &           \makecell[r]{2 (33\%)} &  \makecell[r]{0 (0\%)} &                                                                                                                                                                                                                                                              \makecell[r]{\cite{zhou2020s3}, \cite{kang2021entangled}, \cite{tong2021pattern}, \cite{li2022recguru} \\ } &                                                                               \makecell[r]{\cite{amjadi2021katrec}, \cite{zhang2021behavioral}} &                                                    \makecell[r]{}\\ 
 \midrule 

                  Toys~\cite{he2016ups} &                 5 &            \makecell[r]{0 (0\%)} & \makecell[r]{\textbf{5 (100\%)}} &  \makecell[r]{0 (0\%)} &                                                                                                                                                                                                                                                                                                                                                           \makecell[r]{} & \makecell[r]{\cite{zhou2020s3}, \cite{fan2021continuous}, \cite{li2021lightweight},  \\ \cite{zhang2021behavioral}, \cite{bian2021contrastive}} &                                                          \makecell[r]{} \\

\midrule               
\makecell[l]{\textbf{Total}} & 134 & \makecell[r]{\textbf{86 (64\%)}} & 32 (23 \%) & 16 (12 \%) & & & \\ 
\bottomrule
\end{tabular}
\end{table*}

\section{BERT4Rec implementations}
\label{sec:implementations}
To understand the reasons \srs{for the} poor replicability of BERT4Rec results, demonstrated by our systematic review, we analyse \sre{the} available BERT4Rec implementations. During our systematic review, we identified three available BERT4Rec implementations cited in the papers: %
(1) the original implementation, provided by the authors of the original paper~\cite{sun2019bert4rec}; (2) RecBole~\cite{zhao2021recbole} - a library that contains a large number of recommendation models, including BERT4Rec (3) BERT4Rec-VAE - a project that implements of BERT4Rec and Variational Autoencoder~\cite{liang2018variational} models using PyTorch.\footnote{\href{https://github.com/jaywonchung/BERT4Rec-VAE-Pytorch}{https://github.com/jaywonchung/BERT4Rec-VAE-Pytorch}} 

In addition to these three existing implementations, we \sre{note} the Transformers4Rec project~\cite{de2021transformers4rec}, which is based on the popular Hugging Face Transformers natural language processing library~\cite{wolf2019huggingface}. Transformers4Rec does not include an implementation of BERT4Rec, however, inspired by this project we implement our own version of BERT4Rec that uses Hugging Face's version of BERT as a backbone. \sre{Indeed, given }\srs{the popularity of the Hugging Face Transformers library, we expect that models implemented using it will be highly efficient and  may outperform other available implementations.}%

Salient characteristics of all four BERT4Rec implementations are listed in Table~\ref{table:bert4rec_impl}. We now turn to the experiments with these four implementations, which help us to understand poor replicability of BERT4Rec results.

\begin{table*}[tb]
\caption{BERT4Rec implementations used in our experiments.}
\label{table:bert4rec_impl}
\begin{tabular}{lllrl}
\toprule
Implementation & GitHub URL & Framework & \makecell[l]{GitHub \\ stars}  & \makecell[l]{Example \\ papers} \\
\midrule
Original & \href{https://github.com/FeiSun/BERT4Rec}{FeiSun/BERT4Rec} & Tensorflow v1 & 390 & \cite{hu2021next, huang2021position, liu2021contrastive} \\
RecBole & \href{https://github.com/RUCAIBox/RecBole}{RUCAIBox/RecBole} & PyTorch & 1\scr{,}800 & \cite{fan2021lighter, bian2021contrastive, li2021hyperbolic}\\
BERT4Rec-VAE & \href{https://github.com/jaywonchung/BERT4Rec-VAE-Pytorch}{jaywonchung/BERT4Rec-VAE-Pytorch} & PyTorch & 183 & \cite{rappaz2021recommendation, zeng2021knowledge, tran2022implicit} \\
Ours/Hugging Face & \href{https://github.com/asash/bert4rec_repro}{asash/bert4rec\_repro} & Tensorflow v2 & N/A & N/A \\
\bottomrule
\end{tabular}
\end{table*}

\section{Experimental Setup}
\label{sec:experiments:setup}

\subsection{Research Questions}\label{ssec:expsetup:rq}
We aim to address following research questions with our experiments: 
\begin{enumerate}[font={\bfseries}, label={RQ\arabic*}]
    \item Can we replicate the state-of-the-art results reported in~\cite{sun2019bert4rec} using all of the available BERT4Rec implementations, applying their default configurations?\label{rq:reproduce}
    
    \item \srs{What is the effect of the training time on the performance of the original BERT4Rec implementation?} \label{rq:training_time}
    
    \item Can our Hugging Face-based implementation of BERT4Rec benefit from replacing BERT with another language model available in the Hugging Face Transformers library?\label{rq:transformers}
\end{enumerate}

\subsection{Datasets}
Because our goal is to examine replicability of the BERT4Rec, we use same four datasets as used in the original publication~\cite{sun2019bert4rec}, namely ML-1M~\cite{harper2015movielens},  Beauty~\cite{he2016ups}, Steam~\cite{kang2018sasrec} and ML-20M~\cite{harper2015movielens}. \srs{These datasets also represent 4 of \sre{the} 5 most popular datasets in our systematic review.} \srm{To ensure quality of the data and following \sre{the} common practice~\cite{kang2018sasrec, rendle2010fpmc, tang2018caser, he2017neural}, Sun et al.~\cite{sun2019bert4rec} discarded sequences with less than 5 interactions}\dnote{why or what effect?}. \srm{In order to stay consistent with their experimental setup}, for ML-1M, Steam, and Beauty we use the preprocessed versions of the datasets from the original BERT4Rec repository.\footnote{\href{https://github.com/FeiSun/BERT4Rec/tree/master/data}{https://github.com/FeiSun/BERT4Rec/tree/master/data}} \srs{Table~\ref{table:datasets} \srs{presents} salient characteristics of our experimental datasets.}  %

\sre{We split each dataset into train, validation and test partitions using a Leave-One-Out strategy: for each user, we hold the final interaction for the test set; for a validation set, we select the second last action for 2048 randomly selected users from each dataset; the remaining interactions are used for training.}

\begin{table}[tb]
\caption{Experimental datasets.}
\label{table:datasets}
\begin{tabular}{lrrrrrr}
\toprule
Dataset &  Users &  Items &  Interactions &  Avg. len. &  Sparsity \\
\midrule
ML-1M  &       6\scr{,}040 &       3\scr{,}416 &            999\scr{,}611 &           165.49 &           0.9515 \\
Steam  &     281\scr{,}428 &      13\scr{,}044 &           3\scr{,}488\scr{,}885 &            12.398 &               0.9990 \\
Beauty &      40\scr{,}226 &      54\scr{,}542 &            353\scr{,}962 &             8.79 &                   0.9998 \\
ML-20M          &     138\scr{,}493 &      26\scr{,}744 &          20\scr{,}000\scr{,}263 &           144.41 &   0.9946 \\
\bottomrule
\end{tabular}
\end{table}

\subsection{Models}
\label{ssec:models}
For our replicability experiment (\ref{rq:reproduce}\dnote{why only RQ1 here? or make experiments singular -- made singular}), we use  four BERT4Rec implementations, as described in \sre{S}ection \ref{sec:implementations}.  For all four implementations, we apply their default data prepossessing pipelines and default model parameters. Table~\ref{table:model_config} lists the default parameter values. \sre{The} original BERT4Rec version includes slightly different parameters tuned for each different datasets and we use the ones specified for the ML-20M dataset; however according to the original publication~\cite{sun2019bert4rec} these changes in parameters within the limits specified in the configuration files have only a limited effect and therefore should not change overall model performance\inote{this looks important, should we make a quotation?}. For the ML-1M and ML-20M datasets we also experiment with a  so-called ``longer seq" version, where we increase maximum sequence length, because the average number of items per user is much larger in these two datasets. \scr{Overall, the configuration of our implementation of BERT4Rec is similar to the original model, with a notable difference in the training stopping criteria: to ensure that the models are fully trained, we use an early stopping mechanism; we measure the value of loss function on the validation data and stop training if validation loss did not improve for 200 epochs.}

We also use two baselines in our replicability experiments: (1) MF-BPR~\cite{rendle2009bpr} - a classic matrix factorisation based approach with a pairwise \scr{BPR} loss.  We use the implementation of this model from the LightFM library~\cite{kula2015lightfm} \scr{and set the number of latent components to 128}; (2) SASRec~\cite{kang2018sasrec} - a Transformer-based model, described in Section~\ref{ssec:sasrec}. We use an adaptation of the original code for this model.\scr{\footnote{\href{https://github.com/kang205/SASRec}{https://github.com/kang205/SASRec}}} \scr{For SASRec we set sequence length to 50, embedding size to 50 and use 2 transformer blocks; akkording to Kang et~al.\ these parameters are within the range where SASRec shows reasonable performance.}

In the extra training experiment (\ref{rq:training_time}) we use \sre{the} default configuration of the original BERT4Rec model with the exception of number of training steps, which we vary between 200,000 and 12,800,000. 

Finally, to compare BERT with other models available in the Hugging Face Transformers library (\ref{rq:transformers}), we experiment with two recent Transformer-based architectures: (1) DeBERTa~\cite{he2020deberta} - a model that improves BERT with \sre{a} disentangled attention mechanism~\cite{he2020deberta} where each word is encoded using two vectors (a vector for content and a vector for position); (2) ALBERT~\cite{lan2019albert} - a model that improves BERT via separating the \sre{size} of the hidden state of the vocabulary embedding \sre{from} the number of the hidden layers. It also introduces cross-layer parameter sharing, \sre{which} allows to decrease overall number of parameters in the network.  Following the BERT4Rec naming convention, we call these two models DeBERTa4Rec and ALBERT4Rec. For these two models we use same model configuration parameters as for BERT4Rec with the exception of longer sequence length, which for which we considered values from \{50, 100, 200\} and report the results for the \sre{best-performing} value 200 \scr{according to NDCG@10 metric}.

\subsection{Metrics}\label{ssec:metrics}
Following Sun et al.~\cite{sun2019bert4rec}, \sre{we evaluate the models using the Leave-One-Out strategy} and focus on two ranking-based metrics: NDCG and Recall@K \sre{on the test data}.\footnote{Original BERT4Rec publication uses name Hit Rate (HR) instead of Recall. Hit Rate represent probability of successful prediction of one single interaction in the next-item prediction task. In case of sequential recommendation, when there is only one true positive item per user, this metric is equivalent to Recall, and we prefer more conventional name.} The \sre{authors} report results using \emph{sampled metrics}: for each positive \sre{item} they sample 100 negative \sre{items} - these 101 items are then ranked for evaluation. In particular, they use \sre{a} popularity-based sampling of negatives: the probability of sampling an item as a negative is proportional to its overall popularity in the dataset. However, sampled metrics are known to be problematic. Indeed, it has been shown in a number of recent publications~\cite{rendle2020sampling, canamares2020sampling, dallmann2021case} \sre{that} sampled metrics \sre{are not always consistent with full, unsampled versions (where all items are ranked for each user) and} can lead to incorrect model comparisons. Nevertheless, because our goal is to examine the reproducibility of the BERT4Rec model, we apply sampled metrics in to compare our results with the results reported in the original paper. In addition, following the recommendations in~\cite{rendle2020sampling} and~\cite{canamares2020sampling}, we report results on \emph{unsampled metrics}, where we rank all items from the dataset at evaluation time. However, in experiments where we do not need to compare with the results reported in the original publication~\scr{\cite{sun2019bert4rec}} we only report unsampled metrics. 

\begin{table*}[tb]
\caption{Default parameters of the BERT4Rec implementations.}
\label{table:model_config}
\begin{tabular}{llllll}
\toprule
\textbf{Implementation}    & \textbf{Original} & \textbf{RecBole} & \textbf{BERT4Rec-VAE} & \textbf{Ours}                                                         & \textbf{Ours (longer seq)}                                              \\ \midrule
Sequence length            & 200               & 50               & 100                   & 50                                                                   & 100                                                                   \\
Training stopping criteria & 400\scr{,}000 steps      & 300 epochs       & 200 epochs            & \begin{tabular}[c]{@{}l@{}}Early stopping:\\ 200 epochs\end{tabular} & \begin{tabular}[c]{@{}l@{}}Early stopping: \\ 200 epochs\end{tabular} \\
Item masking probability   & 0.2               & 0.2              & 0.15                  & 0.2                                                                  & 0.2                                                                   \\
Embedding size             & 64                & 64               & 256                   & 64                                                                   & 64                                                                    \\
Transformer blocks         & 2                 & 2                & 2                     & 2                                                                    & 2                                                                     \\
Attention heads            & 2                 & 2                & 4                     & 2                                                                    & 2                                                                     \\ 
\bottomrule
\end{tabular}
\end{table*}

\subsection{Criteria for a Successful Replication}
There is some level of randomness involved in deep neural networks training. Indeed, even when the same model is trained multiple times, random weights initialisation and random data shuffling can lead to small changes in the \sre{resulting effectiveness metrics}. BERT4Rec also randomly masks items during training, which also slightly increases \sre{the} randomness of the process. Furthermore, as mentioned above, the evaluation metrics used by the original BERT4Rec publication~\cite{sun2019bert4rec} involve sampling random items and therefore may change even when evaluating exactly same model.

Thus, it is almost impossible to exactly replicate the reported results. Therefore, we need to define some interval around the values reported in~\cite{sun2019bert4rec} within which we can say that our model replicates originally reported results.  \srs{Because we only know a single measurement per dataset, we can not rely on}  hypothesis testing methods, such as paired t-test, so we need a heuristic for defining the \srs{tolerance} interval. %

According to Madhyastha and Jain~\cite{madhyastha2019model}, the standard deviation in classification metrics for instances of the same deep neural model trained with different random seeds can reach 2.65\%. \sre{If we assume that ranking metrics, such as NDCG and Recall, exhibit similar variance under different model instances, this suggests that a larger tolerance is needed for defining equivalence. Hence}, in this paper we define that a \emph{model replicates the results reported in the original publication on a metric, \srs{if the metric value is equal to the originally reported value within a relative tolerance of $\pm5\%$.}}%

\section{Experimental Results}
\label{sec:experiments:results}
\subsection{\ref{rq:reproduce}. Replicability with default configuration\craig{s}.}
We first analyse if the available implementations of BERT4Rec are able to replicate the originally reported results, when trained with \srs{their} default configuration\craig{s}.
Table~\ref{table:replicability} compares the results achieved by the models trained with \sre{their} default configuration with the results reported in the original BERT4Rec paper~\cite{sun2019bert4rec}. As we can see from the table, both versions of our Hugging Face-based implementation replicate the reported results on 3 datasets out of 4 (all except Beauty) \craig{for} the popularity-sampled Recall@10 metric and on 2 datasets out 4  \craig{for} the popularity-sampled NDCG@10 metric. Moreover, BERT4Rec-VAE replicates the originally reported results on 2 datasets   \craig{for}  the popularity-sampled Recall@10 metrics (both versions of MovieLens) and on one dataset  \craig{for}  NDCG@10 metric (ML-1M). However, the original and RecBole implementations of BERT4Rec fail to replicate \craig{the} originally reported results on all four datasets by a large margin (e.g.\ original implementation is 46.47\% worse according to Recall@10 metric on Steam). 

The table also reports comparison of the BERT4Rec implementations with the \craig{two baseline recommender models}. As we can see, the best version of BERT4Rec always statistically significantly (according to the 2-tailed t-test with Bonferoni multiple test correction, $pvalue < 0.05$) outperforms both MF-BPR and SASRec on all four datasets and on both sampled and unsampled metrics. BERT4Rec-VAE performs better than the baselines on all four datasets, and \dnote{cannot parse. which versions?} \sre{the} \srs{Ours and Ours (longer seq) versions of BERT4Rec} perform better than baselines on three out of four datasets (except Beauty). SASRec performs better than the original and RecBole implementations on three datasets out of four datasets (except Beauty), which \srs{echoes} \dnote{too strong - it "echoes"?} the inconsistencies in our systematic review. \dnote{Further or indeed?} Further, comparisons of the RecBole and original implementations with MF-BPR on Steam and ML-20M can lead to different conclusions if we look to sampled or unsampled metrics. For example according to sampled Recall@10, \craig{the} RecBole implementation is 33\% worse than MF-BPR on Steam dataset, however  at the same time it is 86\% better than MF-BPR according to full Recall@10. This discrepancy is in line with the results reported in~\cite{canamares2020sampling, rendle2020sampling, dallmann2021case} and shows importance of using unsampled metrics in the research. \inote{which formulation do we trust more?}

From the table, we also see that there is big discrepancy in training times.\srs{\footnote{We report training times on our hardware configuration:  16 (out of 32) cores of an AMD Ryzen 3975WX CPU; NVIDIA A6000 GPU;  128GB memory.}} For example, on the ML-1M dataset, training BERT4Rec-VAE takes only 18 minutes, whereas training the RecBole version takes 5.6 hours. At the same time, BERT4Rec-VAE achieves 2.3 times better performance \srs{than RecBole}, according to unsampled Recall@10.

Unfortunately, we fail to replicate results reported in~\cite{sun2019bert4rec} for the Beauty dataset. In attempts to achieve originally reported results, we also ran experiments with the original implementation with a configuration tuned for this specific dataset, tried \sre{training of the} original model for up to 16x more training time, and tried using \sre{the} evaluation framework and metrics from the original codebase. None of these attempts allowed us to reach metric values significantly better than we obtained from the BERT4Rec-VAE implementation, which are 22.68\% worse than the reported values. 

Overall, the answer to the~\ref{rq:reproduce} depends on the implementation and on the dataset. According to popularity-sampled Recall@10, we can replicate \sre{the} originally reported results with our implementation in 3/4 cases, in 2/4 cases using BERT4Rec-VAE and in 0/4 cases using original and RecBole implementations. \sre{Furthermore, for unsampled metrics, we can observe similar conclusions: BERT4Rec models that fail to replicate the  originally reported (sampled) results also performed poorly on unsampled metrics as well.}

\begin{table*}
\centering
\caption{Replicability of the originally reported BERT4Rec results}
\label{table:replicability}
\begin{threeparttable}

\subfloat[ML-1M Dataset]{
\begin{tabular}{|l|l|ll|ll|r|}
\hline
                                                                             & \multirow{2}{*}{Model} & \multicolumn{2}{c|}{Popularity-sampled}                           & \multicolumn{2}{c|}{Unsampled} & \multirow{2}{*}{\begin{tabular}[c]{@{}l@{}}Training \\ Time\end{tabular}} \\ \cline{3-6}
                                                                             &                        & Recall@10                       & NDCG@10                         & Recall@10      & NDCG@10       &                                                                           \\ \hline
\multirow{2}{*}{Baselines}                                                   & MF-BPR                 & 0.5134 (-26.34\%)†              & 0.2736 (-43.21\%)†              & 0.0740†        & 0.0377†       & 58                                                                        \\
                                                                             & SASRec                 & 0.6370 (-8.61\%)†               & 0.4033 (-16.29\%)†              & 0.1993†        & 0.1078†       & 316                                                                       \\ \hline
\multirow{5}{*}{\begin{tabular}[c]{@{}l@{}}BERT4Rec\\ versions\end{tabular}} & Original               & 0.5215 (-25,18\%)†              & 0.3042 (-36.86\%)†              & 0.1518†        & 0.0806†       & 2\scr{,}665                                                                      \\
                                                                             & RecBole                & 0.4562 (-34\scr{.}55\%)†              & 0.2589† (-46.26\%)†             & 0.1061†        & 0.0546†       & 20\scr{,}499                                                                     \\
                                                                             & BERT4Rec-VAE           & \textbf{0.6698 (-3.90\%)†}      & 0.4533 (-5.29\%)†               & 0.2394†        & 0.1314†       & 1\scr{,}085                                                                      \\
                                                                             & Ours                   & \textbf{0.6865 (-1.51\%)}       & \textbf{0.4602 (-4.48\%)}       & 0.2584         & 0.1392        & 3\scr{,}679                                                                      \\
                                                                             & Ours (longer seq)      & {\ul \textbf{0.6975 (+0.07\%)}} & {\ul \textbf{0.4751 (-1.39\%)}} & {\ul 0.2821}   & {\ul 0.1516}  & 2\scr{,}889                                                                      \\ \hline
Reported~\cite{sun2019bert4rec}                                           & BERT4Rec               & 0.6970                          & 0.4818                          & N/A            & N/A           & N/A                                                                       \\ \hline
\end{tabular}
}

\subfloat[Steam Dataset]{
\begin{tabular}{|l|l|ll|ll|r|}
\hline
                                                                             & \multirow{2}{*}{Model} & \multicolumn{2}{c|}{Popularity-sampled}                           & \multicolumn{2}{c|}{Unsampled} & \multirow{2}{*}{\begin{tabular}[c]{@{}l@{}}Training \\ Time\end{tabular}} \\ \cline{3-6}
                                                                             &                        & Recall@10                       & NDCG@10                         & Recall@10      & NDCG@10       &                                                                           \\ \hline
\multirow{2}{*}{Baselines}                                                   & MF-BPR                 & 0.3466 (-13.63\%)†              & 0.1842 (-18.53\%)†              & 0.0398†        & 0.0207†       & 162                                                                       \\
                                                                             & SASRec                 & 0.3744 (-6.70\%)†               & 0.2052 (-9.24\%)†               & 0.1198†        & 0.0482†       & 3\scr{,}614                                                                      \\ \hline
\multirow{4}{*}{\begin{tabular}[c]{@{}l@{}}BERT4Rec\\ versions\end{tabular}} & Original               & 0.2148 (-46.47\%)†              & 0.1064 (-52\scr{.}94\%)†              & 0.0737†        & 0.0375†       & 4\scr{,}847                                                                      \\
                                                                             & RecBole                & 0.2325 (-42.06\%)†              & 0.1177 (-47.94)†                & 0.0744†        & 0.0377†       & 83\scr{,}816                                                                     \\
                                                                             & BERT4Rec-VAE           & 0.3520 (-12.29\%)†              & 0.1941 (-14.15\%)†              & 0.1237†        & 0.0526†       & 65\scr{,}303                                                                     \\
                                                                             & Ours                   & {\ul \textbf{0.3978 (-0.87\%)}} & {\ul \textbf{0.2219 (-1.86\%)}} & {\ul 0.1361}   & {\ul 0.0734}  & 117\scr{,}651                                                                    \\ \hline
Reported~\cite{sun2019bert4rec}                                            & BERT4Rec               & 0.4013                          & 0.2261                          & N/A            & N/A           & N/A                                                                       \\ \hline
\end{tabular}
}

\subfloat[Beauty Dataset]{
\begin{tabular}{|l|l|ll|ll|r|}
\hline
                                                                             & \multirow{2}{*}{Model} & \multicolumn{2}{c|}{Popularity-sampled}           & \multicolumn{2}{c|}{Unsampled} & \multirow{2}{*}{\begin{tabular}[c]{@{}l@{}}Training \\ Time\end{tabular}} \\ \cline{3-6}
                                                                             &                        & Recall@10               & NDCG@10                 & Recall@10      & NDCG@10       &                                                                           \\ \hline
\multirow{2}{*}{Baselines}                                                   & MF-BPR                 & 0.2090 (-30.91\%)†      & 0.1089 (-41\scr{.}47\%)†      & 0.0185†        & 0.0090†       & 58                                                                        \\
                                                                             & SASRec                 & 0.1111 (-63.27\%)†      & 0.0524 (-71\scr{.}83\%)†      & 0.0079†        & 0.0036†       & 316                                                                       \\ \hline
\multirow{4}{*}{\begin{tabular}[c]{@{}l@{}}BERT4Rec\\ versions\end{tabular}} & Original               & 0.1099 (-63.67\%)†      & 0.0567 (-69\scr{.}55\%)†      & 0.0163†        & 0.0079†       & 3\scr{,}249                                                                      \\
                                                                             & RecBole                & 0.1996 (-34.02\%)†      & 0.1103 (-40\scr{.}76\%)†      & 0.0158†        & 0.0079†       & 11\scr{,}024                                                                     \\
                                                                             & BERT4Rec-VAE           & {\ul 0.2339 (-22.68\%)} & {\ul 0.1407 (-24\scr{.}44\%)} & {\ul 0.0331}   & {\ul 0.0188}  & 21\scr{,}426                                                                     \\
                                                                             & Ours                   & 0.1891 (-37\scr{.}49)†        & 0.0919 (-50\scr{.}64\%)†      & 0.0166†        & 0.0080†       & 14\scr{,}497                                                                     \\ \hline
Reported~\cite{sun2019bert4rec}                                             & BERT4Rec               & 0.3025                  & 0.1862                  & N/A            & N/A           & N/A                                                                       \\ \hline
\end{tabular}
}

\subfloat[ML-20M Dataset]{
\begin{tabular}{|l|l|ll|ll|r|}
\hline
                                                                             & \multirow{2}{*}{Model} & \multicolumn{2}{c|}{Popularity-sampled}                           & \multicolumn{2}{c|}{Unsampled} & \multirow{2}{*}{\begin{tabular}[c]{@{}l@{}}Training \\ Time\end{tabular}} \\ \cline{3-6}
                                                                             &                        & Recall@10                       & NDCG@10                         & Recall@10      & NDCG@10       &                                                                           \\ \hline
\multirow{2}{*}{Baselines}                                                   & MF-BPR                 & 0.6126 (-18.02\%)†              & 0.3424 (-35\scr{.}88\%)†              & 0.0807†        & 0.0407†       & 197                                                                       \\
                                                                             & SASRec                 & 0.6582 (-11.92\%)†              & 0.4002 (-25,06\%)†              & 0.1439†        & 0.0724†       & 3635                                                                      \\ \hline
\multirow{5}{*}{\begin{tabular}[c]{@{}l@{}}BERT4Rec\\ versions\end{tabular}} & Original               & 0.4027 (-46.11\%)†              & 0.2193† (-58\scr{.}93\%)†             & 0.0939†        & 0.0474†       & 6\scr{,}029                                                                      \\
                                                                             & RecBole                & 0.4611 (-38.30\%)†              & 0.2589 (-51\scr{.}52\%)†              & 0.0906†         & 0.0753†      & 519\scr{,}666                                                                    \\
                                                                             & BERT4Rec-VAE           & {\ul \textbf{0.7409 (-0.86\%)}} & {\ul \textbf{0.5259 (-1.52\%)}} & {\ul 0.2886}   & {\ul 0.1732}  & 23\scr{,}030                                                                     \\
                                                                             & Ours                   & \textbf{0.7127 (-4.63\%)†}      & 0.4805 (-10.02\%)†              & 0.2393†        & 0.1310†       & 44\scr{,}610                                                                     \\
                                                                             & Ours (longer seq)      & \textbf{0.7268 (-2.74)†}        & 0.4980 (-6.74\%)†               & 0.2514†        & 0.1456†       & 39\scr{,}632                                                                     \\ \hline
Reported~\cite{sun2019bert4rec}                                                               & BERT4Rec               & 0.7473                          & 0.5340                          & N/A            & N/A           & N/A                                                                       \\ \hline
\end{tabular}
}

\begin{tablenotes}
\item \footnotesize \textbf{Bold} denotes a successful replication of a metric reported in~\cite{sun2019bert4rec}, percentages show difference with the reported metric, {\ul underlined} denotes the best model by a metric, † denotes a statistically significant difference with the best model on the paired t-test with Bonferoni multiple testing correction~($pvalue < 0.05$). 
\end{tablenotes}

\end{threeparttable}

\end{table*}

\subsection{\ref{rq:training_time} \srs{Effects of training time on the original BERT4Rec implementation performance}}
\begin{figure*}[tb]
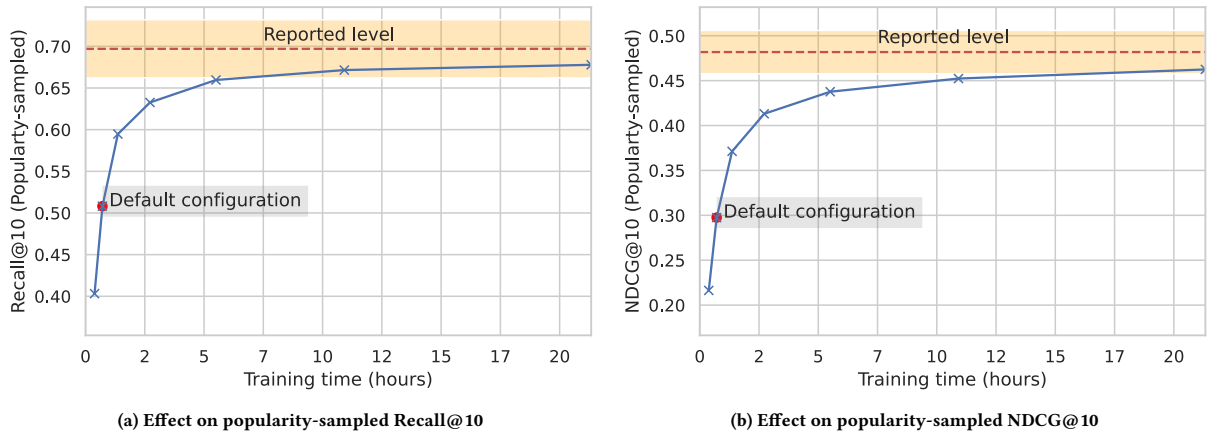

    \subfloat[Effect on popularity-sampled Recall@10]{
        \label{subfig:steps_recall}
        \includesvg[width=0.45\linewidth]{sampled_recall.svg}
    }
    \subfloat[Effect on popularity-sampled NDCG@10]{
        \label{subfig:time_recall}
        \includesvg[width=0.45\linewidth]{sampled_ndcg.svg}
    }
  \caption{Effect of the training time on the performance of the original implementation of BERT4Rec.
  The dashed line represents the metric value reported in~\cite{sun2019bert4rec}, filled area represents $\pm$5\% interval around the reported value, within which we count the result as a "replication" of the originally reported result. Default configuration corresponds to the results with the parameters specified  in the \href{https://github.com/FeiSun/BERT4Rec/blob/master/run_ml-1m.sh}{original repository} for the ML-1M dataset.
  }\label{fig:steps_time_plot}
\end{figure*}

\dnote{reform RQ as discussed?}
\srs{Sun et al.~\cite{sun2019bert4rec} did not report the amount of training they needed to reach the state-of-the-art metrics. The amount of training specified in the original BERT4Rec code is controlled by the parameter \emph{training steps}, which is set to 400,000 by default.}
\srs{To understand the effect of training time, we  vary the number of training steps from slightly less than default (200,000, 0.5x compared to the default) to much larger (1,280,000, 32x compared to the default), which results in a variation of training time from 23 minutes to 21.3 hours. \dnote{what is a and b?} \dnote{why sampled metrics?}}
Figure~\ref{fig:steps_time_plot} portrays the relationship between training time and test performance of the original BERT4Rec implementation on \craig{the} ML-1M dataset \srs{using popularity-sampled metrics.\footnote{We use popularity-sampled metrics, because we need to compare results with the values reported by~Sun et~al.~\cite{sun2019bert4rec}}} As can be observed from the figure, it is possible to reproduce \sre{the} originally reported results using \sre{the} original BERT4Rec implementation, however it requires many times more training time \craig{than the default configuration}. Indeed, the model needs to be trained almost 11 hours instead of 42 minutes to replicate results on sampled Recall@10 (15x times) and 21 hours to replicate results on sampled NDCG@10 (30x times). \srs{\sre{Looking back at} Figure~\ref{fig:main}, the effect of training time on unsampled version of NDCG follows the same general trend: the performance that the model reaches with the default configuration is 51.8\% lower compared to what can be reach with 30x increased amount of training.}

\dnote{what did the original paper say about training time? how would everyone make this mistake?}

Overall \srs{in answer} to~\ref{rq:training_time}, \sre{we} find \srs{that the number of training steps set in the default configuration of the original BERT4Rec code is too small, and to reach performance levels reported by Sun et al.~\cite{sun2019bert4rec}, the training time has to be increased up 30x times (from less than an hour to almost a day of training). This change makes the model much harder to train, and for example, makes hyperparameters tuning much less feasible, specifically with limited amount of hardware. It also makes it very easy to mistakenly use an underfitted version of BERT4Rec in the experiments.} \sre{This likely explains inconsistencies observed in our systematic review.}

\subsection{\ref{rq:transformers} Other Transformers from Hugging Face \srs{for sequential recommendation}}\label{sec:rq3}
To answer our \sre{final} research question, we compare our BERT4Rec implementation with two other models: DeBERTa4Rec (based on the DeBERTa~\cite{lan2019albert} architecture) and ALBERT4Rec (based on \sre{the} ALBERT~\cite{he2020deberta} architecture). \srs{Both architectures propose improvements to \sre{the} BERT architecture (some of the improvements we describe in Section~\ref{ssec:models}), and both models were shown to outperform BERT on \sre{the} GLUE benchmark in their native natural language processing domain. Our goal is to experiment if these improvements hold in the domain of sequential recommendations. Our Hugging Face based implementation makes such an experiment very simple, as in the Transformers library these models have compatible interfaces with BERT. \scr{We also use same early stopping mechanism as we use in our version of BERT4Rec (see Section~\ref{ssec:models}).}}

The results for these \sre{new} models on the ML-1M dataset are listed in Table~\ref{table:other_transformers}. As we can see from the table, \sre{the} both models are better outperform BERT4Rec, however only for ALBERT4Rec are the results  statistically significant (+6.48\% Recall@10, +9.23\% NDCG@10). \dnote{what is our motivation? what is our expectation? do these outperform BERT in other tasks? } These improvements allow us to conclude that the answer to~\ref{rq:transformers} is positive - our implementation of BERT4Rec can further other models available in Hugging Face Transformers library. Overall, there are more than 100 model architectures available in the Transformers library, and many of these models can be used seamlessly instead of BERT with our BERT4Rec implementations.  

\begin{table}[tb]
\caption{Comparison of the Transformer models from the Hugging Face library.  All results are reported on full (unsampled) metrics. The percentage shows the difference with  BERT4Rec. \textbf{Bold} denotes the best model by a metric,† denotes statistically significant difference compared with BERT4Rec according to a paired t-test with Bonferroni multiple testing correction~($pvalue < 0.05$). 
}
\label{table:other_transformers}
\begin{tabular}{llll}
\toprule
Model       & Recall@10                  & NDCG@10                    & \makecell{Training\\Time} \\ \midrule
BERT4Rec    & 0.282                     & 0.151                     & 2\scr{,}889          \\
DeBERTa4Rec & 0.290 (+3.0\%)           & 0.159 (+2.3\%)           & 12\scr{,}114         \\
ALBERT4Rec  & \textbf{0.300 (+6.4\%)†} & \textbf{0.165 (+9.2\%)†} & 12\scr{,}453         \\ 
\bottomrule
\end{tabular}
\end{table}

\section{Related Work \& Discussion}\label{sec:discussion}
As observed in Section~4, there has been some difficulty in replicating the BERT4Rec reported results in~\cite{sun2019bert4rec}. In terms of related work, we highlight Dacrema et al.~\cite{dacrema2019we} who showed that deep learning-based recommendation papers frequently used weak configurations of baselines. They demonstrated this on an example of simpler matrix factorisation baselines, however in this work we show, that this problem also exists with more complicated (and effective) models. Chin et al.~\cite{chin2022datasets} showed that the dataset selection can also influence the conclusion about models performance, which echoes the results of our systematic review. Krichene et al.~\cite{rendle2020sampling} showed that conclusions about model performances may change when sampled or unsampled metrics are used -- this was recently specifically confirmed for sequential recommendation~\cite{dallmann2021case}. Our experiments also confirm these results. However, none of the aforementioned works performed a systematic review about one popular technique, nor examined the reasons why that technique could be difficult to replicate.

Our results for RQ1 and RQ2 in Section~5 above suggest that available BERT4Rec implementations are the cause of this difficulty, probably due to underfitting.
Indeed, some of the recent papers that seemingly have used an underfitted BERT4Rec as a baseline for evaluation on the ML-1M dataset are listed in Table~\ref{table:underfit_papers}. All these papers used unsampled metrics for evaluation, so we can compare their reported values with a fully-trained BERT4Rec results. As we can see from the table, the difference with our fully fitted version of BERT4Rec ranges from -10\% to -53\%, the results that are inline with the ones we observe when trained BERT4Rec implementations with default configurations. Moreover, the results reported for the best models in these papers are rather worse than those we observe for a fully-trained BERT4Rec (GRU4Rec+ -20.59\%, LightSANs -19.03\%) or only marginally better (NOVA-BERT +1.55\%, DuoRec +4.43\%) models. Furthermore, as shown Section~\ref{sec:rq3}, similar or even better improvements can be achieved by a simple replacement of one Transformer with another.  According to this table, ALBERT4Rec improves the state-of-the-art result (+2\% Recall@10 compared to DuoRec), however this improvement needs to be verified via direct comparison with statistical significance testing, rather than comparing reported numbers. \srs{We leave these experiments as well as experiments on more datasets and with other Transformer architectures available in the Hugging Face Transformers Library for future research.}

\begin{table}[tb]
\caption{BERT4Rec results and best model results reported in the literature for the ML-1M dataset. 
The results are copied from the respective publications. All results are reported on full (unsampled) metrics. The percentage shows the difference with our implementation of BERT4Rec. \textbf{Bold} denotes the best reported BERT4Rec result and the best result overall. 
}\label{table:underfit_papers}
\begin{threeparttable}

\begin{tabular}{llll}
\toprule
Publication                              & \begin{tabular}[c]{@{}l@{}}BERT4Rec \\ Recall@10\end{tabular} & Best model & \begin{tabular}[c]{@{}l@{}}Best model \\ Recall@10\end{tabular} \\ \midrule
This paper        & \textbf{0.282 (+0.0\%)}                                     & ALBERT4Rec & \textbf{0.300 (+6.4\%)}                                       \\
Dallmann et al. \cite{dallmann2021case} & 0.160 (-43.2\%)                                              & GRU4Rec+   & 0.224 (-20.5\%)                                                \\
Qiu et al. \cite{qiu2022contrastive}         & 0.132 (-53.1\%)                                             & DuoRec     & 0.294 (+4.4\%)                                                \\
Fan et al. \cite{fan2021lighter}        & 0.221 (-21.6\%)                                             & LightSANs  & 0.228 (-19.0\%)                                               \\
Liu et al. \cite{liu2021noninvasive}    & 0.252 (-10.5\%)                                             & NOVA-BERT  & 0.286 (+1.5\%)                                                \\ \bottomrule
\end{tabular}
\end{threeparttable}

\end{table}

\section{Conclusions}\label{sec:conclusion}
\sre{In this paper we conducted a systematic review of the papers comparing BERT4Rec and SASRec and found that the reported results were not consistent among these papers. To understand the reasons of the inconsistency, we analysed the available BERT4Rec implementations and found that in many cases they fail to replicate originally reported results when trained with their default parameters. Furthermore,  we showed that the original implementation requires much more training time compared to the default configuration in order to replicate originally reported results. This gives weight to the argument that in some cases papers have used underfitted versions of BERT4Rec as baselines.}

\sre{We also proposed our own implementation of BERT4Rec based on the Hugging Face Transformers library, which in most of the cases replicates the originally reported results with the default configuration parameters. We showed that our implementation achieves similar results to the most recent sequential recommendation models, such as NOVA-BERT and DuoRec. We also showed our implementation can be further improved by using other architectures available in the Hugging Face Transformers library. We believe that this paper, as well as our openly available code, will help the researchers to use appropriately trained baselines and will move the science in the right direction.}

\balance
\bibliographystyle{ACM-Reference-Format}
\bibliography{references.bib}

\end{document}